\documentclass[aps,pra,twocolumn,amsmath,nofootinbib]{revtex4-1}
\usepackage{epsfig}
\usepackage[colorlinks,linkcolor=blue,anchorcolor=blue,
            citecolor=blue,urlcolor=blue,
            breaklinks=true]{hyperref}
\usepackage{graphics}
\usepackage{color}
\usepackage{graphicx}
\usepackage{dcolumn}
\usepackage{bm}
\usepackage{cases}
\usepackage{appendix}
\usepackage{ulem}

\newcommand{\tb}{\textbf}

\begin{document}
\author{Yong-Long Wang$^{1,2,3}$}
 \email{Email: wangyonglong@lyu.edu.cn}
\author{Hua Jiang$^{3}$}
\author{Hong-Shi Zong$^{4,5,6}$}
\email{Email: zonghs@nju.edu.cn}
\address{$^{1}$ National Laboratory of Solid State Microstructures, Department of Materials Science and Engineering, Nanjing University, Nanjing 210093, China}
\address{$^{2}$ Collaborative Innovation Center of Advanced Microstructures, Nanjing University, Nanjing 210093, China}
\address{$^{3}$ School of Physics and Electronic Engineering, Linyi University, Linyi 276005, China}
\address{$^{4}$ Department of Physics, Nanjing University, Nanjing 210093, China}
\address{$^{5}$ Joint Center for Particle, Nuclear Physics and Cosmology, Nanjing 210093, China}
\address{$^{6}$ State Key Laboratory of Theoretical Physics, Institute of Theoretical Physics, CAS, Beijing 100190, China}

\title{Geometric influences of a particle confined to a curved surface embedded in three-dimensional Euclidean space}
\begin{abstract}
In the spirit of the thin-layer quantization approach, we give the formula of the geometric influences of a particle confined to a curved surface embedded in three-dimensional Euclidean space. The geometric contributions can result from the reduced commutation relation between the acted function depending on normal variable and the normal derivative. According to the formula, we obtain the geometric potential, geometric momentum, geometric orbital angular momentum, geometric linear Rashba and cubic Dresselhaus spin-orbit couplings. As an example, a truncated cone surface is considered. We find that the geometric orbital angular momentum can provide an azimuthal polarization for spin, and the sign of the geometric Dresselhaus spin-orbit coupling can be flipped through the inclination angle of generatrix.
\bigskip

\end{abstract}
\maketitle

\section{INTRODUCTION}\label{1}
The remarkable development of nanotechnology has initiated experimental insights into the geometric influences on the motion on a curved surface embedded in three-dimensional (3D) Euclidean space~\cite{Bowick2009, Turner2010,Streubel2016}. An important contribution is the geometric influence on kinetic energy, the so-called geometric potential~\cite{HJensen1971, Costa1981, Wang2016}. The effective potential has been realized  experimentally in photonic topological crystal~\cite{Szameit2010}, and the geometric influence on the quantum transport of two-dimensional (2D) curved materials has been investigated~\cite{Santos2016, Wang2016JPD, Amorim2016}. Another important result is the geometric influence on momentum~\cite{Liu2007, Liu2011} that has been observed governing the propagation of surface plasmon on metallic wires~\cite{Spittel2015}. For full generality, as a curved surface is embedded in a higher-dimensional (HD) Euclidean space, a novel geometrically induced gauge potential is present only when the space of normal states is degenerate~\cite{Jaffe2003}. For a spinless charged particle confined to a space curve, the geometric gauge potential is identical to a magnetic moment in the presence of electromagnetic field~\cite{Brandt2015}. All of these results have enriched the confining potential theory.

While the geometric influence on quantum transport of matter is becoming a hot topic in condensed matter, the topic of magnetism in curved geometry is evolving into an independent research field of modern magnetism with many exciting theoretical predictions and strong application potential ~\cite{Streubel2015, Streubel2016}. Under the circumstances, the researching of the geometric influences on orbital angular momentum (OAM) and on spin-orbit coupling (SOC)~\cite{Jeong2011, Chang2013, Ortix2015, Shikakhwa2016} becomes very necessary.

For a particle confined to a curved surface, the thin-layer quantization approach (TLQA) has been longstanding discussion in quantum mechanics~\cite{Costa1981, Kaplan1997, Wang2016}. After two decades, the method was extended to a case containing external electromagnetic field~\cite{Encinosa2006, Ferrari2008, BJensen2009, Ortix2011}. In the presence of external electromagnetic field, the performing sequence plays an essential role in the validity of the TLQA~\cite{Wang2016}. The sequence is determined by two final aims of the TLQA. One is to separate surface quantum equation from normal component analytically. The other is to preserve the information about the normal motion in the effective Hamiltonian as much as possible. The former defines the valid conditions of the TLQA~\cite{Costa1981, BJensen2009}. The latter determines the performing sequence~\cite{Wang2016}. Although the quantization procedure has been developed very well, the explicit formula of geometric influences still needs further investigation, especially when a physical operator contains terms of high-order momentum operator. As an example, for spin-$1/2$ particles confined to a curved surface the geometric influences are complicated considering the linear Rashba and cubic Dresselhaus spin-orbit interactions~\cite{Ortix2015}. Therefore, the formula of geometric influences is useful to simplify the TLQA and to extend the potential of applications.

In the present paper, we will study the geometric influence on a physical operator which depends on normal derivative. In Sec. II, the formula of geometric influence is given for a $\partial_3$-dependent physical operator. By using the formula, the geometric potential, geometric momentum, geometric orbital angular momentum, geometric linear Rashba and cubic Dresselhaus spin-orbit couplings are obtained. In Sec. III, with a truncated cone being an example, the geometric influences are calculated for Hamiltonian, momentum, orbital angular momentum, linear Rashba spin-orbit coupling and cubic Dresselhaus spin-orbit coupling. Finally, our conclusions and further thoughts are contained in Sec. IV.

\section{The formula of geometric influence of a particle confined to a curved surface}\label{2}
In quantum mechanics, the state of a microscopic particle can be described by a wave function, and the physical operator with respect to an observable is Hermitian. In the same vein, in  curvilinear coordinate system (CCS) a momentum operator can be expressed by the derivative with respect to the associated curvilinear coordinate variable. For a particle constrained to a curved surface, the TLQA can achieve the separation of surface equation and normal component by introducing a confining potential in normal direction. The confining potential raises the energy of normal excitations far beyond the energy scale associated with motion tangent to the surface. It is straightforward to determine that the motion in the normal direction lies in the ground state of the confining potential. Naturally, the commutation relation between normal variable and normal momentum is held in the ground state. As their incompatibility does not depend on normal variable, the effect of the normal commutation relation is preserved in the effective dynamics as an additive geometric influence due to the confinement in normal direction.

Describing a particle confined to a curved surface, the wave function $\psi$ is a function of $q_1$, $q_2$ and $q_3$, and the density probability in a volume element $d\tau$ is $|\psi|^2d\tau$ that trivially satisfies $\int|\psi|^2d\tau=1$, which can be expanded as
\begin{equation}\label{Equality}
\begin{split}
\int|\psi|^2d\tau&=\int|\psi|^2\sqrt{G}dq_1dq_2dq_3\\
& =\int|\psi|^2f\sqrt{g}dq_1dq_2dq_3=1.
\end{split}
\end{equation}
Here $G$ and $g$ are two determinants with respect to the two metric $G_{ij}(i,j=1,2,3)$ and $g_{ab}(a,b=1,2)$, respectively, defined in Appendix A, and the rescaled factor $f$ is usually a function of $q_1$, $q_2$ and $q_3$. In normal direction, the particle lies in the ground state of the confining potential. Naturally, the wave function of the normal ground state can stand for the normal component of the wave function $\psi$. Absorbing the rescaled factor $f$ from the volume element, a new wave function $\chi$ can be expressed as
\begin{equation}\label{WFTrans}
\chi=\sqrt{f}\psi\quad {\rm{or}} \quad \psi=\frac{1}{\sqrt{f}}\chi.
\end{equation}
The best advantage of the new function $\chi$ is that it can be analytically separated into a surface component $\chi_s$ and a normal component $\chi_n$, $\chi(q_1, q_2, q_3)=\chi_s(q_1, q_2)\chi_n(q_3)$.

Without losing generality, we consider an arbitrary Hermitian operator $\hat{\tb{F}}$ that can be completely expressed by momentum operator and coordinate variable. From the previous discussions, we know that once the operator $\hat{\tb{F}}$ acts on the wave function $\psi$, it will act on the factor $\frac{1}{\sqrt{f}}$, too. In terms of the expression of $f$ (see in Appendix A), it is easy to prove that the two components of momentum operator on the surface and the three curvilinear coordinate variables are all commutative with $\frac{1}{\sqrt{f}}$ under the protection of limiting $q_3\to 0$. However, the normal component $-i\hbar\partial_3$ does not commute with $\frac{1}{\sqrt{f}}$, which provides a geometric influence on the effective operator. In other words, the geometric influence can be determined by the commutation relation between $-i\hbar\partial_3$ and $\frac{1}{\sqrt{f}}$ with $q_3\to 0$.

For convenience, the operator $\hat{\tb{F}}$ can be divided into two components,
\begin{equation}\label{Operator}
\hat{\tb{F}}=\hat{\tb{F}}_0+\hat{\tb{F}}^{\prime},
\end{equation}
where $\hat{\tb{F}}_0$ does not depend on $\partial_3$, but $\hat{\tb{F}}^{\prime}$ does. Furthermore, $\hat{\tb{F}}^{\prime}$ can be redivided into two subcomponents,
\begin{equation}\label{POperator}
\hat{\tb{F}}^{\prime}=\hat{\tb{F}}_0^{\prime}+\hat{\tb{F}}_n^{\prime},
\end{equation}
where the operator $\hat{\tb{F}}_0^{\prime}$ consists only of the terms in which $\partial_3$ operator automatically vanishes in the calculating procedure, and $\hat{\tb{F}}_n^{\prime}$ contains the others.

In light of the previously mentioned discussions, it is easy to obtain
\begin{equation}\label{TrivialRelation}
[\hat{F}_0,\frac{1}{\sqrt{f}}]_0=0,
\end{equation}
where $[\cdot,\cdot]_0=\lim_{q_3\to0}[\cdot,\cdot]$, $[\cdot,\cdot]$ denotes the usual Dirac bracket. For convenience, $[\cdot,\cdot]_0$ is named as a reduced Dirac bracket and its associated commutation relation as a reduced commutation relation (RCR). Using the reduced Dirac bracket, the geometric influence of $\hat{\tb{F}}$ determined by $\frac{1}{\sqrt{f}}$ can be expressed as
\begin{equation}\label{GOperatorP}
\hat{\tb{F}}_{ng}=[\hat{\tb{F}}^{\prime}_n,\frac{1}{\sqrt{f}}]_0.
\end{equation}
The geometric influence $\hat{\tb{F}}_{ng}$ completely results from the $\partial_3$ dependence of $\hat{\tb{F}}_n^{\prime}$ and the $q_3$ dependence of $\frac{1}{\sqrt{f}}$. In other words, $\hat{\tb{F}}_{ng}$ originates from the RCR between $\partial_3$ and $q_3$. Initially, the operator $\hat{\tb{F}}_0^{\prime}$ does depend on $\partial_3$ operator. Therefore, the geometric influence on $\hat{\tb{F}}$ is given not only by $\hat{\tb{F}}_{n}^{\prime}$, but also by $\hat{\tb{F}}_0^{\prime}$, which can be briefly expressed as
\begin{equation}\label{GOperator}
\hat{\tb{F}}_g=\hat{\tb{F}}_{0g}+\hat{\tb{F}}_{ng} =[\hat{\tb{O}}(\partial_3), f(\cdot,q_3)]_0,
\end{equation}
where $\hat{\tb{F}}_{0g}=\lim_{q_3\to 0}\hat{\tb{F}}_0^{\prime}$, $\hat{\tb{O}}(\partial_3)$ stands for a variety of $\partial_3$-dependent factors that appear in $\hat{\tb{F}}^{\prime}$, $f(\cdot,q_3)$ stands for $\frac{1}{\sqrt{f}}$ and a variety of $q_3$-dependent factors which are acted by $\partial_3$ in calculating $\hat{\tb{F}}^{\prime}$, wherein "$\cdot$" denotes $q_1$ and $q_2$. In the calculation procedure, a rule needs to be clarified: in the operator $\hat{\tb{F}}^{\prime}$, $\partial_3$ contributes $[\partial_3, f(\cdot,q_3)]_0$, $(\partial_3)^2$ does $[\partial_3,[\partial_3, f(\cdot,q_3)]]_0$, and so on. For the operator $(\partial_3)^n$, its geometric contribution can be defined by
\begin{equation}\nonumber
[\partial_3,\underbrace{[\partial_3, [\cdots[\partial_3,f(\cdot, q_3)]]]}_{n-1}]_0.
\end{equation}
The rule is protected by the fundamental framework of the TLQA in~\cite{Wang2016}.

As a consequence, the geometric influences on the motion on the curved surface embedded in 3D Euclidean space can be described by the RCR between $\partial_3$ and the acted $q_3$-dependent function, because that $f(\cdot,q_3)$ and $\underbrace{[\partial_3,[\cdots,f(\cdot,q_3)]]}_i \quad (i=1,\cdots,n-1)$ are all functions of $q_3$, otherwise their geometric influences will vanish. Eq.~\eqref{GOperator} is the central result of this paper, a formula of geometric influence. With the formula, for the final effective operator $\hat{\tb{F}}_E$ the rest work is to obtain the surface component that can be defined by
\begin{equation}\label{sOperator}
\hat{\tb{F}}_s=\lim_{q_3\to0}\hat{\tb{F}}_0,
\end{equation}
and then the effective operator $\hat{\tb{F}}_E$ can be
\begin{equation}\label{EOperator}
\hat{\tb{F}}_E=\hat{\tb{F}}_s+\hat{\tb{F}}_g.
\end{equation}

In what follows, we will calculate the effective Hamiltonian, effective momentum, effective orbit angular momentum, effective linear Rashba spin-orbit coupling and effective cubic Dresselhaus spin-orbit coupling for particles confined to a curved surface. The Hamiltonian of a free particle is
\begin{equation}\label{Hamilton}
\tb{H}=-\frac{\hbar^2}{2m}\frac{1}{\sqrt{G}}\partial_i(\sqrt{G}G^{ij}\partial_j),
\end{equation}
where $\hbar$ is the Planck constant divided by $2\pi$, $m$ is the mass of the particle, $\partial_i$ is the derivative with respect to $q_i$, the index values are $1$, $2$ and $3$, $G$ and $G^{ij}$ are the determinant and inverse of the metric $G_{ij}$ that is defined in Appendix A. According to Eqs.~\eqref{Operator}, ~\eqref{POperator}, ~\eqref{GOperator}, ~\eqref{sOperator} and ~\eqref{EOperator}, the effective Hamiltonian $\tb{H}_E$ can be obtained
\begin{equation}\label{EHamilton}
\tb{H}_E=-\frac{\hbar^2}{2m}\frac{1}{\sqrt{g}}\partial_a(\sqrt{g}g^{ab}\partial_b) -\frac{\hbar^2}{2m}(\rm{M}^2-\rm{K}),
\end{equation}
where $\rm{M}$ is the mean curvature, $\rm{K}$ is the Gaussian curvature, $g$ and $g^{ab}$ are the determinant and inverse of the reduced metric $g_{ab}$ ($a, b=1, 2$) that is defined in Appendix A. On the right hand side of Eq.~\eqref{EHamilton}, the first term is the surface Hamiltonian, the second is the so-called geometric potential $\tb{V}_g$ which is provided by the normal Hamiltonian and the rescaled factor $f$ together. Specifically, the geometric potential is given by the RCRs between $\partial_3$ and $\frac{1}{\sqrt{f}}$, $\partial_3$ and $f$, and $(\partial_3)^2$ and $\frac{1}{\sqrt{f}}$.

In the same case, the momentum operator $\vec{\tb{P}}$ is
\begin{equation}\label{Momentum}
\vec{\tb{P}}=-i\hbar(\vec{\tb{e}}_1\frac{1}{\sqrt{G_{11}}}\partial_1 +\vec{\tb{e}}_2\frac{1}{\sqrt{G_{22}}}\partial_2 +\vec{\tb{e}}_n\frac{1}{\sqrt{G_{33}}}\partial_3),
\end{equation}
where $\vec{\tb{e}}_1$, $\vec{\tb{e}}_2$ and $\vec{\tb{e}}_n$ are three unit vectors with respect to $q_1$, $q_2$ and $q_3$, respectively. In terms of Eqs.~\eqref{Operator} and ~\eqref{GOperator}, the geometric momentum can be obtained
\begin{equation}\label{GMomentum}
\vec{\tb{P}}_g=i\hbar{\rm{M}}\vec{\tb{e}}_n,
\end{equation}
where $\rm{M}$ is the mean curvature. The geometric momentum is determined by the RCR between $\partial_3$ and $\frac{1}{\sqrt{f}}$. The definition of Eq.~\eqref{GMomentum} without the surface component is different from the original geometric momentum~\cite{Liu2007, Liu2011} which is identical to our effective momentum. Our effective momentum consists of surface and geometric components, that is
\begin{equation}\label{EMomentum}
\vec{\tb{P}}_E=-i\hbar(\vec{\tb{e}}_1\frac{1}{\sqrt{g_{11}}}\partial_1 +\vec{\tb{e}}_2\frac{1}{\sqrt{g_{22}}}\partial_2) +\vec{\tb{P}}_g,
\end{equation}
where the terms in the round bracket denote the surface momentum and $\vec{\tb{P}}_g$ is the geometric momentum.

Furthermore, with the expression of the effective momentum Eq.~\eqref{EMomentum} and the definition of an orbital angular momentum $\vec{\tb{L}}=\vec{\tb{r}}\times\vec{\tb{P}}$, the effective OAM can be obtained
\begin{equation}\label{EAMomentum}
\begin{split}
\vec{\tb{L}}_E=-i\hbar[&(\vec{\tb{r}}\times\vec{\tb{e}}_1) \frac{1}{\sqrt{g_{11}}}\partial_1+(\vec{\tb{r}}\times\vec{\tb{e}}_2) \frac{1}{\sqrt{g_{22}}}\partial_2]+\vec{\tb{L}}_g.
\end{split}
\end{equation}
On the right hand side, the two terms in square bracket stand for the surface OAM and $\vec{\tb{L}}_g$ is the geometric OAM, that is
\begin{equation}\label{GAMomentum}
\vec{\tb{L}}_g=i\hbar(\vec{\tb{r}}\times \vec{\tb{e}}_n)\rm{M}.
\end{equation}

In recent years, the manipulation of spin transport is achieved by using the effective magnetic field due to the spin-orbit interaction. Linear Rashba and cubic Dresselhaus SOCs are usually employed to control the spin transport in 2D materials. The different behaviors of SOCs can be defined by different curved structures, and thus the physical effects of curved materials have been widely studied~\cite{Kuemmeth2008, Levy2010, Steele2013, Laird2015}. For a curved surface with large curvature, both Rashba and Dresselhaus SOCs will play an important role in controlling spin transport. In terms of Eqs.~\eqref{Operator}, ~\eqref{POperator}, ~\eqref{GOperatorP}, ~\eqref{GOperator}, ~\eqref{sOperator} and ~\eqref{EOperator}, we can discuss the exact expressions of the linear Rashba and cubic Dresselhaus SOCs on a curved surface.

Before the discussion, we first introduce some relevant constants. In CCS, the Pauli matrices $\sigma^i$, Rashba tensor $S_{ij}$ and Dresselhaus tensor $S_{iijj}$ can be defined by~\cite{Chang2013}
\begin{equation}\label{TensorTrans}
\begin{split}
& \sigma^i=\frac{\partial q^i}{\partial x^s}\sigma^s,\\
& S_{ij}=\frac{\partial x^s}{\partial q^i}\frac{\partial x^t}{\partial q^j}S_{st},\\
& S_{iijj}=\frac{\partial x^s}{\partial q^i}\frac{\partial x^s}{\partial q^i} \frac{\partial x^t}{\partial q^j} \frac{\partial x^t}{\partial q^j}S_{sstt},
\end{split}
\end{equation}
where $\sigma^s$, $S_{st}$ and $S_{sstt}$ are the associated Pauli matrices, Rashba tensor and Dresselhaus tensor in Cartesian coordinate system, respectively. The index $i$ and $j$ values are $1$, $2$ and $3$, standing for three curvilinear coordinate variables. The index $s$ and $t$ values are also $1$, $2$ and $3$, denoting $x$, $y$ and $z$. The repeated indices indicate the Einstein summation convention throughout the present paper. For simplicity, for a $[111]$-grown quantum well the components of the Rashba tensor~\cite{Chang2013} are taken as
\begin{equation}\label{RDTensor}
\begin{split}
& S_{xy}=S_{yz}=S_{zx}=\frac{\alpha}{\hbar},\\
& S_{zy}=S_{yx}=S_{xz}=-\frac{\alpha}{\hbar},
\end{split}
\end{equation}
and for a $[100]$-grown quantum well the components of the Dresselhaus tensor~\cite{Chang2013} are
\begin{equation}\label{DTensor}
\begin{split}
& S_{xxyy}=S_{yyzz}=S_{zzxx}=\frac{\beta}{\hbar^3},\\
& S_{zzyy}=S_{yyxx}=S_{xxzz}=-\frac{\beta}{\hbar^3},
\end{split}
\end{equation}
where $\alpha$ is the Rashba coupling strength whose unit is $\rm{meVnm}$, and $\beta$ is the Dresselhaus coupling strength whose unit is $\rm{meVnm^3}$.

The linear Rashba SOC is
\begin{equation}\label{RashbaSOC}
\tb{H}^{R}=-i\hbar S_{ij}\sigma^i G^{jk}\partial_k,
\end{equation}
where $i\neq j$, $S_{ij}$ describe the components of Rashba tensor, $\sigma^i$ stand for Pauli matrices, $G^{jk}$ are the contravariant components of the metric $G_{jk}$, $\partial_k$ are the derivatives with respect to $q_k$, and the index values are $1$, $2$ and $3$. From Eqs.~\eqref{Operator} and ~\eqref{GOperator}, we can obtain the geometric Rashba SOC as
\begin{equation}\label{GRSOC}
\tb{H}^R_g=i\hbar S_{i3}^0\sigma^i_0 \rm{M},
\end{equation}
where $i=1,2,3$, $S_{i3}^0$ are the components of the reduced Rashba tensor (defined in Appendix B), $\sigma^i_0$ are the reduced Pauli matrices (defined in Appendix B) and $\rm{M}$ is the mean curvature. This geometric Rashba SOC depending on spin can be given by the RCR between $\partial_3$ and $\frac{1}{\sqrt{f}}$. Subsequently, the effective Rashba SOC can be expressed as
\begin{equation}\label{ERSOC}
\tb{H}^{R}_E=-i\hbar S_{ia}^0\sigma^i_0 g^{ab}\partial_b+\tb{H}^R_g,
\end{equation}
where $a,b=1,2$, $S_{ia}^0$ are the components of the reduced Rashba tensor (defined in Appendix B), and $g^{ab}$ are the contravariant components of the reduced metric $g_{ab}$. The first term is the surface Rashba SOC. For a system with a large curvature, the geometric Rashba SOC can induce a large difference in the behavior of spin electrons. Practically, the manipulation of spin transport can be achieved by designing the geometry of nanodevices.

The cubic Dresselhaus SOC is
\begin{equation}\label{DSOC}
\tb{H}^{D}
=i\hbar^3S_{iijj}\sigma^i G^{ik}\partial_k[G^{jl}\partial_l (G^{jm}\partial_m)],
\end{equation}
where $i\neq j$, $S_{iijj}$ denote the components of Dresselhaus tensor, $G^{ik}$ is the inverse of the metric $G_{ij}$, $\partial_k$ are the derivatives with respect to $q_k$, and the index values are $1$, $2$ and $3$. From Eqs.~\eqref{Operator} and ~\eqref{GOperator}, we can also obtain the geometric Dresselhaus SOC as
\begin{equation}\label{GDSOC}
\begin{split}
\tb{H}^D_g=& i\hbar^3S_{33aa}^0\sigma^3_0g^{ab}_1 \partial_b(g^{ac}\partial_c)\\
& -i\hbar^3S_{33aa}^0\sigma^3_0g^{ab} \partial_b(g^{ab}_1\partial_c)\\
&-i\hbar^3 S_{33aa}^0\sigma^3_0(g^{ab}\partial_b(g^{ac} \partial_c)){\rm{M}}\\
& +i\hbar^3 S_{aa33}^0\sigma^a_0g^{ab}\partial_b (3{\rm{M}}^2-{\rm{K}}),
\end{split}
\end{equation}
where $a,b,c=1,2$, $S_{33aa}^0$ and $S_{aa33}^0$ are the components of the reduced Dresselhaus tensor (defined in Appendix B), $\sigma_0^3$ and $\sigma_0^a$ are the reduced Pauli matrices (defined in Appendix B), $g^{ab}$ and $g^{ac}$ are the contravariant components of the reduced metric $g_{ab}$, $g_1^{ab}$ are the components determined by the RCR between $\partial_3$ and $G^{ab}$, $g^{ab}_1=[\partial_3, G^{ab}]_0$, $\rm{M}$ is the mean curvature, and $\rm{K}$ is the Gaussian curvature. On the right hand side of Eq.~\eqref{GDSOC}, the first and second terms are given by the terms in Eq.~\eqref{DSOC} corresponding to $\hat{\tb{F}}_0^{\prime}$ in Eq.~\eqref{POperator}, the third and fourth are provided by the ones corresponding to $\hat{\tb{F}}_n^{\prime}$. Naturally, the effective Dresselhaus SOC can be obtained
\begin{equation}\label{EDSOC}
\tb{H}^{D}_{E}=i\hbar^3 S_{aabb}^0\sigma^a_0g^{ac}\partial_c[g^{bd}\partial_b (g^{be}\partial_e)]+\tb{H}_g^D,
\end{equation}
where $a,b,c,d,e=1,2$, $S_{aabb}^0$ are the components of the reduced Dresselhaus tensor (defined in Appendix B). On the right hand side of Eq.~\eqref{EDSOC}, the first term is the surface Dresselhaus SOC, and the second $\tb{H}^D_g$ is the geometric Dresselhaus SOC which can play an important role in controlling the spin transport in 2D curved materials with large curvature.

\section{a truncated cone}\label{4}
\begin{figure}[htbp]
\centering
\includegraphics[scale=0.27]{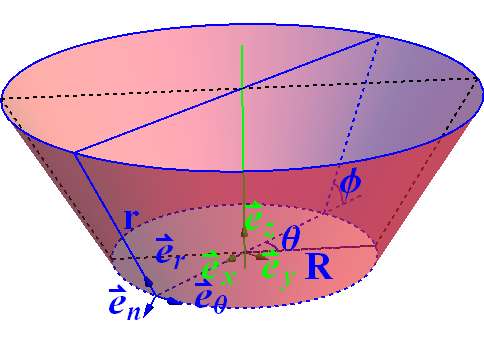}\\
\caption{A truncated cone surface and notations.}\label{Cone}
\end{figure}

As an example, the side surface of a truncated cone is considered~\cite{Navia2005, Silva2012, Gaididei2014}; the minimum radius is $R$ and the length of the generatrix is $l$, shown in Fig.~\ref{Cone}. By varying the generatrix inclination angle $0\leq\phi\leq\frac{\pi}{2}$, one can continuously proceed from a planar ring $(\phi=0)$ to a cylindrical surface $(\phi=\frac{\pi}{2})$. The considered surface can be parametrized by
\begin{equation}\label{ConeSurface}
\vec{\tb{r}}=(w\cos\theta, w\sin\theta, r\sin\phi),
\end{equation}
where $w=R+r\cos\phi$, $\theta$ and $r$ are two curvilinear coordinate variables, $0\leq\theta<2\pi$ and $0\leq r\leq l$.

According to Eqs.~\eqref{EHamilton}, ~\eqref{C-Dg}, ~\eqref{C-Igab}, ~\eqref{C-CGP}, the effective Hamiltonian for a particle confined to the truncated cone surface can be obtained:
\begin{equation}\label{ConeEH}
\begin{split}
H_E(\theta,r)=& -\frac{\hbar^2}{2m}\frac{1}{w^2}\partial_{\theta}^2 -\frac{\hbar^2}{2m}\partial_r^2-\frac{\hbar^2}{2m}\frac{\cos\phi}{w}\partial_r\\
& -\frac{\hbar^2}{8m}\frac{\sin^2\phi}{w^2}.
\end{split}
\end{equation}
On the right-hand side of Eq.~\eqref{ConeEH}, the first term is the $\theta$ component of the kinetic energy, the second and third are the $r$ component, and the fourth is the so-called geometric potential which is the compensation of reducing the normal dimension. Obviously, when $\phi=\frac{\pi}{2}$, there are $\cos\phi=0$ and $\sin\phi=1$, and then the effective Hamiltonian Eq.~\eqref{ConeEH} is naturally identical to that of a cylindrical surface~\cite{Chang2013}.

\begin{figure}[htbp]
\centering
\includegraphics[scale=0.27]{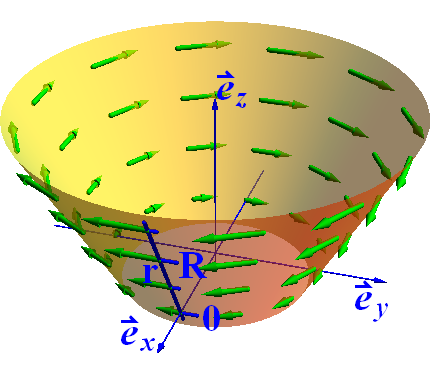}\\
\caption{Schematic of the geometric OAM of $r=0, 1/2R, R, 3/2R$ with $i\hbar R=1$. The length of the green arrows describes the strength of polarization.}\label{OAM}
\end{figure}

In the same case, from Eqs.~\eqref{EMomentum}, ~\eqref{B-GM} and ~\eqref{C-SMetricTensor}, the effective momentum can be obtained:
\begin{equation}\label{ConeEM}
\vec{\tb{P}}_E=-i\hbar\vec{\tb{e}}_{\theta}\frac{1}{w}\partial_{\theta} - i\hbar\vec{\tb{e}}_r\partial_r+ i\hbar\vec{\tb{e}}_n\frac{\sin\phi}{2w}.
\end{equation}
The first term is the $\theta$ component of momentum, the second is the $r$ component, and the third is the geometric momentum $\vec{\tb{P}}_g$ that is contributed by the RCR between $\partial_3$ and $\frac{1}{\sqrt{f}}$. The presence of the geometric momentum destroys the commutation relation among the velocity operators~\cite{Silva2015}. And then the effective OAM on the truncated cone surface can be expressed as
\begin{equation}\label{ConeEAM}
\begin{split}
\vec{\tb{L}}_E=& i\hbar(\vec{\tb{e}}_n \frac{R\cos\phi+r}{w}\partial_{\theta} -\vec{\tb{e}}_r\frac{R\sin\phi}{w}\partial_{\theta})\\
& +i\hbar\vec{\tb{e}}_{\theta}R\sin\phi\partial_r +\vec{\tb{L}}_g.
\end{split}
\end{equation}
On the right-hand side of Eq.~\eqref{ConeEAM}, the first two terms in the round bracket are defined by the $\theta$ component of momentum, the third is determined by the $r$ component of momentum, and the last one is the geometric OAM denoted by $\vec{\tb{L}}_g$, that is
\begin{equation}\label{ConeGAM}
\vec{\tb{L}}_g=i\hbar\vec{\tb{e}}_{\theta}\frac{R\cos\phi+r}{2w}\sin\phi.
\end{equation}
Obviously, the geometric OAM is in the negative $\theta$ direction. For a charged particle with spin moving on the truncated cone surface, the geometric OAM plays the role of magnetic moment which can polarize spin azimuthally. The effect is sketched in Fig.~\ref{OAM}. When the longitudinal component is considered, the effective OAM ~\eqref{ConeEAM} can polarize spin helically~\cite{Streubel2016}.

On the truncated cone surface, the geometric influence also plays an important role in the linear Rashba SOC and cubic Dresselhaus SOC. With the reduced Pauli matrices Eq.~\eqref{C-RPMatrix} and the reduced Rashba tensor Eq.~\eqref{C-RRTensor}, from Eq.~\eqref{ERSOC} we can obtain the effective Rashba SOC as
\begin{widetext}
\begin{equation}\label{ConeRSOC}
\begin{split}
H^{R}_E=& -i\alpha[\cos\theta\sigma^x +\sin\theta \sigma^y-(\sin\theta+\cos\theta)\sigma^z]\frac{1}{w} \partial_{\theta}\\
&+i\alpha[(\sin\phi-\cos\phi\sin\theta)\sigma^x -(\sin\phi-\cos\phi\cos\theta)\sigma^y -\cos\phi(\cos\theta-\sin\theta)\sigma^z]\partial_r\\
&+\frac{1}{2}i\alpha[(\sin^2\phi\sin\theta-\frac{1}{2}\sin2\phi)\sigma^x-(\sin^2\phi \cos\theta+\frac{1}{2}\sin2\phi)\sigma^y-\sin^2\phi(\sin\theta-\cos\theta)\sigma^z] \frac{1}{R}.
\end{split}
\end{equation}
On the right-hand side of Eq.~\eqref{ConeRSOC}, the first-row terms are the $\theta$ component, the second-row terms denote the $r$ component, and the third-row terms stand for the geometric influence on SOC which are determined by the RCR between $\partial_3$ and $\frac{1}{\sqrt{f}}$. When $\phi=\frac{\pi}{2}$, there are $\cos\phi=0$, $\sin\phi=1$ and $\sin2\phi=0$, and thus the effective Rashba SOC can be simplified as
\begin{equation}\label{CylRSOC}
\begin{split}
H^{R}_E=& -i\alpha[\cos\theta\sigma^x+\sin\theta\sigma^y -(\sin\theta+\cos\theta)\sigma^z]\frac{1}{R}\partial_{\theta} +i\alpha(\sigma^x -\sigma^y)\partial_r\\
& +\frac{1}{2}i\alpha[\sin\theta\sigma^x-\cos\theta\sigma^y -(\sin\theta-\cos\theta)\sigma^z]\frac{1}{R}.
\end{split}
\end{equation}
This result is in good agreement with that in~\cite{Chang2013}.

In terms of the reduced Pauli matrices Eq.~\eqref{C-RPMatrix} and the reduced Dresselhaus tensor Eq.~\eqref{C-RDTensor}, from Eq.~\eqref{EDSOC} we can calculate the effective Dresselhaus SOC as
\begin{equation}\label{ConeDSOC}
\begin{split}
H^{D}_E=& i\beta\cos2\phi\cos2\theta[(\cos\phi\cos\theta\sigma^x +\cos\phi\sin\theta\sigma^y+\sin\phi\sigma^z) \frac{1}{w^2}\partial_{\theta}^2\partial_r
+(\sin\theta\sigma^x-\cos\theta\sigma^y) \frac{1}{w}\partial_{\theta}\partial_r^2]\\
& -4i\beta\cos2\phi\cos2\theta\cos\phi (\sin\phi\cos\theta\sigma^x +\sin\phi\sin\theta\sigma^y-\cos\phi\sigma^z)\frac{1}{w}\frac{1}{w^2} \partial_{\theta}^2\\
& +\frac{1}{2}i\beta\cos2\phi\cos2\theta\sin\phi(\sin\phi\cos\theta\sigma^x +\sin\phi\sin\theta\sigma^y-\cos\phi\sigma^z)\frac{1}{w}(\frac{1}{w^2} \partial_{\theta}^2-\partial_r^2)\\
& +\frac{3}{4}i\beta\cos2\phi\cos2\theta\sin^2\phi(-\sin\theta\sigma^x +\cos\theta\sigma^y)\frac{1}{w^3}\partial_{\theta}\\
& +i\beta\cos2\phi\cos2\theta\sin\phi[\frac{1}{4}\sin\phi\cos\phi\cos\theta\sigma^x +\frac{1}{4}\sin\phi\cos\phi\sin\theta\sigma^y+(\frac{1}{4}\sin^2\phi-1)\sigma^z] \frac{1}{w^2}\partial_r\\
&+i\beta\cos2\phi\cos2\theta[\frac{1}{2}\sin^2\phi\cos^2\phi\cos\theta\sigma^x +\frac{1}{2}\sin^2\phi\cos^2\phi\sin\theta\sigma^y+\sin\phi\cos\phi (1+\frac{1}{2}\sin^2\phi)\sigma^z]\frac{1}{w^3}.
\end{split}
\end{equation}
The first-row terms describe the cubic momenta in both unconfined $\theta$ and $r$ directions. The second-row terms are contributed by the RCR between $\partial_3$ and $G^{ab}$. The third-row terms stand for the square momenta of both unconfined $\theta$ and $r$ directions coupling to the geometric momentum provided by $[\partial_3, \frac{1}{\sqrt{f}}]_0$. The fourth- and fifth-row terms are the geometric potential, which is determined by the square momenta confined to the truncated cone surface coupling to the momentum in $\theta$ direction and in $r$ direction, respectively. The sixth-row terms are completely produced by the confinement in the normal direction. It is interesting that $\cos 2\phi$ appears in Eq.~\eqref{ConeDSOC}. Therefore, the sign of the Dresselhaus SOC can be flipped by varying the generatrix inclination angle. Namely, when $0<\phi<\frac{\pi}{4}$, the spin is polarized by the Dresselhaus SOC in a certain direction. However when $\frac{\pi}{4}<\phi<\frac{\pi}{2}$, the spin is polarized in an opposite direction.

When $\phi=\frac{\pi}{2}$, there are $\sin\phi=1$, $\cos\phi=0$, $\sin2\phi=0$ and $\cos2\phi=-1$, and then the effective Dresselhaus SOC Eq.~\eqref{ConeDSOC} is simplified as
\begin{equation}\label{CylDSOC}
\begin{split}
H_E^{D}=& i\beta\cos2\theta[(-\sin\theta\sigma^x+\cos\theta\sigma^y) \frac{1}{R}\partial_{\theta}\partial_r^2-\sigma^z\partial_r(\frac{1}{R^2} \partial_{\theta}^2)]\\
&-\frac{1}{2}i\beta\cos2\theta(\cos\theta\sigma^x+\sin\theta\sigma^y)\frac{1}{R} (\frac{1}{R^2}\partial_{\theta}^2-\partial_r^2)
+\frac{3}{4R^2}i\beta\cos2\theta[(\sin\theta\sigma^x -\cos\theta\sigma^y)\frac{1}{R}\partial_{\theta}
+\sigma^z\partial_r].
\end{split}
\end{equation}
The simplified SOC describes the effective Dresselhaus SOC on a cylindrical surface. This result is identical to the result in ~\cite{Chang2013}.
\end{widetext}

\section{Conclusions and discussions}
In this paper, we studied a particle confined to a curved surface embedded in 3D Euclidean space, and found that the geometric influences can be briefly determined by the RCR between the normal derivative $\partial_3$ and the acted $q_3$-dependent function. This formula provides a shortcut to obtain the geometric influence on an arbitrary $\partial_3$-dependent operator. Using the brief formula, it is easy to calculate the geometric potential, geometric momentum, geometric OAM, and geometric Rashba and Dresselhaus SOCs. As an illustration, a truncated cone surface was considered, and the geometric influences on Hamiltonian, momentum, OAM, linear Rashba SOC and cubic Dresselhaus SOC were discussed specifically. These results show that the geometric influences are considerable effects on a 2D system with large curvature. For instance, the geometric OAM, geometric Rashba SOC and geometric Dresselhaus SOC can be used to manipulate its spin transport. In addition, these discussions are helpful to understand the TLQA entirely, to understand further the quantum properties of 2D curved system embedded in 3D Euclidean space, and are significant to simplify the calculation of geometric influences, especially as a physical operator involving higher order of normal momentum operator.

Attention is necessarily paid to the fact that the present paper implies an assumption that all the geometric influences can be described by the associated operator and the rescaled factor. It is easy to check that the assumption is right as a system by reducing the dimension by 1. However, when the number of reducing dimensions becomes 2 or greater, the confining potential will particularly enrich the geometric influence, such as the torsion-induced geometric potential~\cite{Ortix2015}, geometrically induced gauge potential~\cite{Jaffe2003} and geometrically induced magnetic moment~\cite{Brandt2015}. In the complicated case, the geometric influences can not be described only by the associated operator and the rescaled factor. At the same time, the symmetries of the confining potential have to be considered~\cite{Jaffe2003}. In other words, there are abundant degrees of freedom in the space of the ground states of the confining potential~\cite{Schmelcher2014}, and the associated motion may be preserved in the effective dynamics. The topic needs further investigation. Inspired by the discussions in twisted tubes~\cite{Jaffe1999, Schmelcher2014}, we are investigating the geometric influences on the motion on a space curve embedded in three-dimensional Euclidean space and the details will be reported in another paper soon.

\section*{Acknowledgments}
We thank Fan Wang for fruitful discussions, and we acknowledge the financial support of the National Basic Research Program of China (Grants No. 2013CB632904 and No. 2013CB632702), and the National Nature Science Foundation of China (Grants No. 11625418, No. 11690030, No. 11474158, No. 51472114, No. 11475085, and No. 11535005). Y.-L. W. was funded by Linyi University (LYDX2016BS135).

\section*{Appendix A: Metric Tensor}
This Appendix mainly presents the derivations of the rescaled factor $f$ in Eq.~\eqref{WFTrans}. We assume that a curved surface can be parametrized by
\renewcommand\theequation{A1}
\begin{equation}\label{A-S}
\vec{\tb{r}}=\vec{\tb{r}}(q_1,q_2),
\end{equation}
where $q_1, q_2$ denote two tangential variables in CCS. With respect to $q_1$ and $q_2$, the two unit vectors $\vec{\tb{e}}_1$ and $\vec{\tb{e}}_2$ can be defined by
\renewcommand\theequation{A2}
\begin{equation}\label{A-TangentVector}
\vec{\tb{e}}_1=\frac{\partial_1\vec{\tb{r}}}{|\partial_1\vec{\tb{r}}|},\quad
\vec{\tb{e}}_2=\frac{\partial_2\vec{\tb{r}}}{|\partial_2\vec{\tb{r}}|},
\end{equation}
respectively. Here $\partial_1=\partial/\partial q^1$, $\partial_2=\partial/\partial q^2$, wherein $q^1$ and $q^2$ are contravariant variables with respect to covariant variables $q_1$ and $q_2$, respectively. In terms of $\vec{\tb{e}}_1$ and $\vec{\tb{e}}_2$, the unit vector along the direction normal to the surface Eq.~\eqref{A-S} can be defined by ~\cite{Ono2009}
\renewcommand\theequation{A3}
\begin{equation}\label{A-NormalVector}
\vec{\tb{e}}_n=\frac{\partial_1\vec{\tb{r}}\times\partial_2\vec{\tb{r}}} {|\partial_1\vec{\tb{r}}\times\partial_2\vec{\tb{r}}|}.
\end{equation}
Subsequently, the points near the surface Eq. \eqref{A-S} can be parametrized by
\renewcommand\theequation{A4}
\begin{equation}\label{A-Vn}
\vec{\tb{R}}(q_1,q_2,q_3)=\vec{\tb{r}}(q_1,q_2)+q_3\vec{\tb{e}}_n,
\end{equation}
where $q_3$ is the curvilinear coordinate variable normal to the surface Eq. ~\eqref{A-S}.

From Eq. \eqref{A-Vn}, the covariant components of the metric can be defined by
\renewcommand\theequation{A5}
\begin{equation}\label{A-VnMetricTensor}
G_{ij}=\partial_i\vec{\tb{R}}\cdot \partial_j\vec{\tb{R}},
\end{equation}
where the index values are $1$, $2$ and $3$. By introducing new indices $(a,b=1,2)$, from Eq. \eqref{A-S} the covariant components of the reduced metric can be determined by
\renewcommand\theequation{A6}
\begin{equation}\label{A-SMetricTensor}
g_{ab}=\partial_a\vec{\tb{r}}\cdot \partial_b\vec{\tb{r}}.
\end{equation}
It is straightforward to demonstrate that $G_{ab}$ and $g_{ab}$ satisfy the following relationship
\renewcommand\theequation{A7}
\begin{equation}\label{A-RelationsVnS}
G_{ab}=g_{ab}+(\alpha g+g^T\alpha^T)_{ab}q_3+(\alpha g\alpha^T)_{ab}(q_3)^2,
\end{equation}
and $G_{a3}=G_{3a}=0$, $G_{33}=1$, where $T$ denotes the matrix transpose, $\alpha$ is the Weingarten curvature matrix, the associated elements are defined by
\renewcommand\theequation{A8}
\begin{equation}\label{A-Alpha}
\alpha_{ab}=\frac{1}{g}
\left (
\begin{array}{cc}
g_{12}h_{21}-g_{22}h_{11} & g_{21}h_{11}-g_{11}h_{21}\\
g_{12}h_{22}-g_{22}h_{12} & g_{12}h_{21}-g_{11}h_{22}
\end{array}
\right ),
\end{equation}
wherein $h_{ab}$ are the coefficients of the second fundamental form with the definition $h_{ab}=\vec{\tb{e}}_n\cdot\partial_a\partial_b\vec{\tb{r}}$. Furthermore, it is easy to prove that $G$ and $g$ satisfy the following relation
\renewcommand\theequation{A9}
\begin{equation}\label{A-RelationVnS}
G=f^2g,
\end{equation}
where $G=\det(G_{ij})$, $g=\det(g_{ab})$, and the rescaled factor $f$ is
\renewcommand\theequation{A10}
\begin{equation}\label{A-Factor}
f=1+2{\rm{M}}q_3+{\rm{K}}(q_3)^2,
\end{equation}
wherein ${\rm{M}}=\frac{1}{2}\mathrm{Tr}(\alpha)$ is the mean curvature and $\rm{K}=\det(\alpha)$ is the Gaussian curvature.

\section*{Appendix B: Geometric operators}
For convenience, a reduced Dirac bracket is introduced by $[\cdot, \cdot]_0=\lim_{q_3\to 0}[\cdot, \cdot]$. Using the new bracket and limiting $q_3\to 0$, from Eqs.~\eqref{A-RelationsVnS} and ~\eqref{A-Factor} one can obtain
\renewcommand\theequation{B1}
\begin{equation}\label{B-Limits}
\lim_{q_3\to 0}G^{ab}=g^{ab},\quad \lim_{q_3\to 0}f=1,
\end{equation}
\renewcommand\theequation{B2}
\begin{equation}\label{B-CRDf}
\begin{split}
& \lim_{q_3\to 0}(\partial_3 f)=[\partial_3, f]_0=2\rm{M},\\
& \lim_{q_3\to 0}(\partial_3^2f)=[\partial_3, [\partial_3, f]]_0=2\rm{K},
\end{split}
\end{equation}
and
\renewcommand\theequation{B3}
\begin{equation}\label{B-CRDRf}
\begin{split}
& \lim_{q_3\to0}(\partial_3\frac{1}{\sqrt{f}})=[\partial_3, \frac{1}{\sqrt{f}}]_0=-{\rm{M}},\\
& \lim_{q_3\to 0}(\partial_3^2\frac{1}{\sqrt{f}})=[\partial_3,[\partial_3, \frac{1}{\sqrt{f}}]]_0=3\rm{M}^2-\rm{K},
\end{split}
\end{equation}
where $G^{ab}$ are the contravariant components of the metric $G_{ab}$, $g^{ab}$ are the contravariant components of the reduced metric $g_{ab}$, $\rm{M}$ is the mean curvature, and $\rm{K}$ is the Gaussian curvature.

In view of the dependence of $\partial_3$, the Hamiltonian Eq.~\eqref{Hamilton} can be expanded into
\renewcommand\theequation{B4}
\begin{equation}\label{B-Hamilton}
\begin{split}
\tb{H} & =-\frac{\hbar^2}{2m}\frac{1}{\sqrt{G}}\partial_a(\sqrt{G}G^{ab}\partial_b)\\
&\quad -\frac{\hbar^2}{2m}\frac{1}{\sqrt{G}}\partial_3(\sqrt{G}G^{33}\partial_3)\\
&=\tb{H}_0+\tb{H}^{\prime},
\end{split}
\end{equation}
where $\tb{H}_0$ denotes the terms independent of $\partial_3$
\renewcommand\theequation{B5}
\begin{equation}
\tb{H}_0=-\frac{\hbar^2}{2m}\frac{1}{\sqrt{G}}\partial_a(\sqrt{G}G^{ab}\partial_b),
\end{equation}
and $\tb{H}^{\prime}$ depends on $\partial_3$ can be expanded as
\renewcommand\theequation{B6}
\begin{equation}\label{B-HamiltonP}
\begin{split}
\tb{H}^{\prime}& =-\frac{\hbar^2}{2m}\frac{1}{\sqrt{G}}\partial_3 (\sqrt{G}G^{33}\partial_3)\\
& =-\frac{\hbar^2}{2m}\frac{1}{f\sqrt{g}}\partial_3 (f\sqrt{g}\partial_3)\\
&=-\frac{\hbar^2}{2m}\frac{1}{f}(\partial_3 f)\partial_3 -\frac{\hbar^2}{2m}\partial_3\partial_3,
\end{split}
\end{equation}
here $G^{33}=1$ is considered. Using the reduced Dirac bracket, according to Eqs.~\eqref{GOperatorP}, ~\eqref{B-CRDf}, ~\eqref{B-CRDRf} and ~\eqref{B-Hamilton}, the geometric potential $\tb{V}_g$ can be deduced
\renewcommand\theequation{B7}
\begin{equation}\label{B-GP}
\begin{split}
\tb{V}_g&=\lim_{q_3\to 0}(\tb{H}^{\prime}\frac{1}{\sqrt{f}})\\
&=-\frac{\hbar^2}{2m}\frac{1}{\sqrt{g}}[\frac{1}{f\sqrt{g}}[\partial_3,f] \sqrt{g}\partial_3,\frac{1}{\sqrt{f}}]_0\\
&\quad -\frac{\hbar^2}{2m}[\partial_3, [\partial_3,\frac{1}{\sqrt{f}}]]_0\\
&=-\frac{\hbar^2}{2m}(-2{\rm{M}}^2)-\frac{\hbar^2}{2m}(3{\rm{M}}^2-{\rm{K}})\\
&=-\frac{\hbar^2}{2m}({\rm{M}}^2-{\rm{K}}),
\end{split}
\end{equation}
the result perfectly agrees with that in ~\cite{Costa1981}. In order to obtain the effective Hamiltonian Eq.~\eqref{EHamilton}, from Eq. ~\eqref{B-Limits} the surface Hamiltonian $\tb{H}_s$ is
\renewcommand\theequation{B8}
\begin{equation}\label{B-sHamilton}
\begin{split}
\tb{H}_s& =\lim_{q_3\to 0}\tb{H}_0\\
&=-\frac{\hbar^2}{2m}\frac{1}{\sqrt{g}}\partial_a(\sqrt{g}g^{ab}\partial_b).
\end{split}
\end{equation}

In terms of the dependence of $\partial_3$, the momentum Eq.~\eqref{Momentum} can be divided into two parts $\vec{\tb{P}}_0$ and $\vec{\tb{P}}^{\prime}$. The two components can be expressed as
\renewcommand\theequation{B9}
\begin{equation}\label{B-Momentum0}
\vec{\tb{P}}_0=-i\hbar(\vec{\tb{e}}_1\frac{1}{\sqrt{G_{11}}}\partial_1 +\vec{\tb{e}}_2\frac{1}{\sqrt{G_{22}}}\partial_2),
\end{equation}
and
\renewcommand\theequation{B10}
\begin{equation}\label{B-MomentumP}
\vec{\tb{P}}^{\prime}=-i\hbar(\vec{\tb{e}}_n\frac{1}{\sqrt{G_{33}}}\partial_3),
\end{equation}
respectively. From Eqs.~\eqref{GOperatorP}, ~\eqref{B-CRDRf} and ~\eqref{B-MomentumP}, the geometric momentum is obtained,
\renewcommand\theequation{B11}
\begin{equation}\label{B-GM}
\begin{split}
\vec{\tb{P}}_g&=\lim_{q_3\to 0}(\vec{\tb{P}}^{\prime}\frac{1}{\sqrt{f}})\\
&=-i\hbar\vec{\tb{e}}_n[\partial_3, \frac{1}{\sqrt{f}}]_0\\
&=i\hbar{\rm{M}}\vec{\tb{e}}_n,
\end{split}
\end{equation}
where ${\rm{M}}$ is the mean curvature. By limiting $q_3\to 0$, the surface momentum $\vec{\tb{P}}_s$ can be obtained,
\renewcommand\theequation{B12}
\begin{equation}
\vec{\tb{P}}_s=-i\hbar(\vec{\tb{e}}_1\frac{1}{\sqrt{g_{11}}}\partial_1 +\vec{\tb{e}}_2\frac{1}{\sqrt{g_{22}}}\partial_2).
\end{equation}

Considering the dependence of $\partial_3$, one can divide the Rashba SOC into two components $\tb{H}^{R}_0$ and $\tb{H}^{R\prime}$. The former $\tb{H}^{R}_0$ independent of $\partial_3$ reads
\renewcommand\theequation{B13}
\begin{equation}\label{B-RSOC0}
\tb{H}^{R}_0=-i\hbar S_{ia}\sigma^iG^{ab}\partial_b,
\end{equation}
and the latter $\tb{H}^{R\prime}$ dependent on $\partial_3$ is
\renewcommand\theequation{B14}
\begin{equation}\label{B-RSOCP}
\tb{H}^{R\prime}=-i\hbar S_{a3}\sigma^a G^{33}\partial_3.
\end{equation}
According to Eqs.~\eqref{GOperatorP}, ~\eqref{B-CRDRf} and ~\eqref{B-RSOCP}, using the reduced Dirac bracket one can obtain the geometric Rashba SOC as
\renewcommand\theequation{B15}
\begin{equation}\label{B-GRSOC}
\begin{split}
\tb{H}^{R}_g&=\lim_{q_3\to0}(\tb{H}^{R\prime}\frac{1}{\sqrt{f}})\\
& =-i\hbar S_{i3}^0\sigma^i_0[\partial_3,\frac{1}{\sqrt{f}}]_0\\
&=i\hbar S_{i3}^0\sigma^i_0{\rm{M}},
\end{split}
\end{equation}
where ${\rm{M}}$ is the mean curvature, $S_{i3}^0$ are reduced components of the Rashba tensor defined by $S_{i3}^0=\lim_{q_3\to 0}S_{i3}$, and $\sigma_0^i$ are reduced Pauli matrices defined by $\sigma_0^i=\lim_{q_3\to 0}\sigma^i$. By limiting $q_3\to 0$, from Eq.~\eqref{B-RSOC0} the surface Rashba SOC can be obtained:
\renewcommand\theequation{B16}
\begin{equation}\label{B-sRSOC}
\begin{split}
\tb{H}_s^{R}&=\lim_{q_3\to 0}\tb{H}^{R}_0\\
& =-i\hbar S_{ia}^0\sigma^i_0 g^{ab}\partial_b,
\end{split}
\end{equation}

In order to study the geometric influences, the Dresselhaus SOC is divided into two parts $\tb{H}^{D}_0$ and $\tb{H}^{D\prime}$. Here $\tb{H}^{D}_0$ independent of $\partial_3$ reads
\renewcommand\theequation{B17}
\begin{equation}\label{B-DSOC0}
\tb{H}^{D}_0=i\hbar^3 S_{aabb}\sigma^a G^{ac}\partial_c[G^{bd}\partial_d (G^{be}\partial_e)],
\end{equation}
$\tb{H}^{D\prime}$ dependent on $\partial_3$ is
\renewcommand\theequation{B18}
\begin{equation}\label{B-DSOCP}
\begin{split}
\tb{H}^{D\prime}=& i\hbar^3 S_{33aa}\partial_3[G^{ab}\partial_b (G^{ac}\partial_c)]\\
& +i\hbar^3 S_{aa33}G^{ab}\partial_b\partial_3^2,
\end{split}
\end{equation}
where $G^{a3}=G^{3a}=0$ and $G^{33}=1$ have been considered. According to the vanishing of $\partial_3$ in calculating process, $\tb{H}^{D\prime}$ can be further divided into two subparts $\tb{H}^{D\prime}_{0}$ and $\tb{H}^{D\prime}_{n}$. They are
\renewcommand\theequation{B19}
\begin{equation}\label{B-DSOCP0}
\begin{split}
\tb{H}^{D\prime}_0=& i\hbar^3 S_{33aa}\sigma^3(\partial_3 G^{ab})\partial_b(G^{ac}\partial_c)\\
& +i\hbar^3 S_{33aa}\sigma^3G^{ab}\partial_b(\partial_3 G^{ac})\partial_c,
\end{split}
\end{equation}
and
\renewcommand\theequation{B20}
\begin{equation}\label{B-DSOCPn}
\begin{split}
\tb{H}^{D\prime}_n=&i\hbar^3 S_{33aa}\sigma^3 G^{ab}\partial_b(G^{ac}\partial_c)\partial_3\\
& +i\hbar^3 S_{aa33}\sigma^a G^{ab}\partial_b\partial_3^2,
\end{split}
\end{equation}
respectively. By limiting $q_3\to 0$, the geometric influence provided by $\tb{H}^{D\prime}_0$ can be obtained,
\renewcommand\theequation{B21}
\begin{equation}\label{B-DSOC0g}
\begin{split}
\tb{H}^{D}_{0g}&=\lim_{q_3\to 0}\tb{H}^{D\prime}_{0}\\
&=i\hbar^3S_{33aa}^0\sigma^3_0g^{ab}_1\partial_b(g^{ac}\partial_c)\\
& \quad +i\hbar^3S_{33aa}^0\sigma^3_0g^{ab}\partial_b(g^{ab}_1\partial_c),
\end{split}
\end{equation}
where $g^{ab}$ are contravariant components of the reduced metric, $S_{33aa}^0$ are components of the reduced Dresselhaus tensor, $\sigma^3_0$ is a reduced Pauli matrix, and $g^{ab}_1$ are defined by
\renewcommand\theequation{B22}
\begin{equation}\label{B-Limits}
\begin{split}
& S_{33aa}^0=\lim_{q_3\to 0}S_{33aa},\\
& \sigma_0^3=\lim_{q_3\to0}\sigma^3,\\
& g_1^{ab}=[\partial_3, G^{ab}]_0.
\end{split}
\end{equation}
Using the reduced Dirac bracket, one can obtain the geometric influence determined by the emergence of $\frac{1}{\sqrt{f}}$ as
\renewcommand\theequation{B23}
\begin{equation}\label{DSCOng}
\begin{split}
\tb{H}^{D}_{ng}&=\lim_{q_3\to 0}(\tb{H}^{D\prime}_n\frac{1}{\sqrt{f}})\\
&=i\hbar^3 S_{33aa}^0\sigma^3_0(g^{ab}\partial_b(g^{ac}\partial_c)) [\partial_3,\frac{1}{\sqrt{f}}]_0\\
& \quad +i\hbar^3 S_{aa33}^0\sigma^a_0g^{ab}\partial_b[\partial_3, [\partial_3,\frac{1}{\sqrt{f}}]]_0\\
&=-i\hbar^3 S_{33aa}^0\sigma^3_0(g^{ab}\partial_b(g^{ac}\partial_c)){\rm{M}}\\
& \quad +i\hbar^3 S_{aa33}^0\sigma^a_0g^{ab}\partial_b(3{\rm{M}}^2-{\rm{K}}),
\end{split}
\end{equation}
where $S_{33aa}^0$ and $S_{aa33}^0$ are the components of the reduced Dresselhaus tensor, $\sigma_0^3$ and $\sigma_0^a$ are the reduced Pauli matrices, $g^{ab}$ are the contravariant components of the reduced metric, ${\rm{M}}$ is the mean curvature, and ${\rm{K}}$ is the Gaussian curvature. It is worthy to notice that ${\rm{M}}$ and ${\rm{K}}$ must be placed after the derivatives with respect to $q_1$ or $q_2$, because ${\rm{M}}$ and ${\rm{K}}$ are usually functions of $q_1$ and $q_2$, and not commute with $\partial_{2,3}$ operators. By virtue of these results, the geometric Dresselhaus SOC can be given by
\renewcommand\theequation{B24}
\begin{equation}\label{B-GDSOC}
\begin{split}
\tb{H}^{D}_g& =\tb{H}^{D}_{0g}+\tb{H}^{D}_{ng}\\
&=i\hbar^3S_{33aa}^0\sigma^3_0g^{ab}_1\partial_b(g^{ac}\partial_c)\\
& \quad -i\hbar^3S_{33aa}^0\sigma^3_0g^{ab}\partial_b(g^{ab}_1\partial_c)\\
&\quad -i\hbar^3 S_{33aa}^0\sigma^3_0(g^{ab}\partial_b(g^{ac}\partial_c)){\rm{M}}\\
& \quad +i\hbar^3 S_{aa33}^0\sigma^a_0g^{ab}\partial_b(3{\rm{M}}^2-{\rm{K}}).
\end{split}
\end{equation}
In the same case, the surface Dresselhaus SOC can be obtained:
\renewcommand\theequation{B25}
\begin{equation}\label{B-sDSOC}
\begin{split}
\tb{H}^{D}_s&=\lim_{q_3\to 0}H^{D}_0\\
&=i\hbar^3 S_{aabb}^0\sigma^a_0 g^{ac}\partial_c[g^{bd}\partial_b(g^{be}\partial_e)]
\end{split}
\end{equation}
with
\renewcommand\theequation{B26}
\begin{equation}\label{B-Limits2}
S_{aabb}^0=\lim_{q_3\to 0}S_{aabb}, \quad \sigma^a_0=\lim_{q_3\to 0}\sigma^a.
\end{equation}
\section*{Appendix C: The geometric parameters of the truncated cone surface}
According to Eqs.~\eqref{ConeSurface},~\eqref{A-TangentVector} and ~\eqref{A-NormalVector}, the two tangent unit vectors $\vec{\tb{e}}_{\theta}$ and $\vec{\tb{e}}_{r}$ on the truncated cone surface can be obtained,
\renewcommand\theequation{C1}
\begin{equation}\label{C-TVector1}
\begin{split}
& \vec{\tb{e}}_{\theta}=(-\sin\theta,\cos\theta,0),\\
& \vec{\tb{e}}_{r}=(\cos\phi\cos\theta,\cos\phi\sin\theta,\sin\phi),
\end{split}
\end{equation}
and the normal unit vector $\vec{\tb{e}}_n$ can be
\renewcommand\theequation{C2}
\begin{equation}\label{C-NVector}
\vec{\tb{e}}_n=(\sin\phi\cos\theta,\sin\phi\sin\theta,-\cos\phi).
\end{equation}
In terms of Eqs.~\eqref{ConeSurface}, ~\eqref{A-Vn} and ~\eqref{C-NVector}, the 3D subspace associated to the truncated cone surface can be parametrized by
\renewcommand\theequation{C3}
\begin{equation}\label{C-Vn}
\vec{\tb{R}}(\theta, r, q_3)=(W\cos\theta, W\sin\theta, r\sin\phi-q_3\cos\phi),
\end{equation}
where
\renewcommand\theequation{C4}
\begin{equation}\label{C-Ww}
W=R+r\cos\phi+q_3\sin\phi,\quad w=R+r\cos\phi.
\end{equation}
From the truncated cone surface Eq. ~\eqref{ConeSurface} and the subspace Eq.~\eqref{C-Vn}, with the definitions of Eqs. ~\eqref{A-VnMetricTensor} and ~\eqref{A-SMetricTensor}, one can obtain
\renewcommand\theequation{C5}
\begin{equation}\label{C-VnMetricTensor}
(G_{ij})=\left (
\begin{array}{ccc}
W^2 & 0 & 0\\
0 & 1 & 0\\
0 & 0 & 1
\end{array}
\right )
\end{equation}
and
\renewcommand\theequation{C6}
\begin{equation}\label{C-SMetricTensor}
(g_{ab})=\left (
\begin{array}{cc}
w^2 & 0\\
0 & 1
\end{array}
\right ),
\end{equation}
respectively. It is straightforward to obtain
\renewcommand\theequation{C7}
\begin{equation}\label{C-Dg}
g=\det(g_{ab})=w^2,
\end{equation}
and
\renewcommand\theequation{C8}
\begin{equation}\label{C-DG}
G=gf^2,
\end{equation}
where the rescaled factor $f$ is
\renewcommand\theequation{C9}
\begin{equation}\label{C-f}
f=1+\frac{\sin\phi}{w}q_3.
\end{equation}
Comparing Eq.~\eqref{C-f} with Eq.~\eqref{A-Factor}, one can obtain the mean curvature $\rm{M}$ and Gaussian curvature $\rm{K}$ as
\renewcommand\theequation{C10}
\begin{equation}\label{C-MeanGaussian}
{\rm{M}}=\frac{\sin\phi}{2w},\quad {\rm{K}}=0,
\end{equation}
respectively. From Eqs.~\eqref{C-VnMetricTensor} and ~\eqref{C-SMetricTensor}, one can obtain the inverse matrices $(g^{ab})$ and $(G^{ab})$ as
\renewcommand\theequation{C11}
\begin{equation}\label{C-Igab}
(g^{ab})=
\left (
\begin{array}{cc}
\frac{1}{w^2} & 0\\
0 & 1
\end{array}
\right ),
\end{equation}
and
\renewcommand\theequation{C12}
\begin{equation}\label{C-IGab}
(G^{ab})=
\left (
\begin{array}{cc}
\frac{1}{W^2} & 0\\
0 & 1\\
\end{array}
\right ),\\
\end{equation}
respectively. By using the reduced Dirac bracket, from Eq.~\eqref{B-Limits} $(g^{ab}_1)$ is obtained,
\renewcommand\theequation{C13}
\begin{equation}\label{C-IG1ab}
(g_1^{ab})=[\partial_3, G^{ab}]_0=
\left (
\begin{array}{cc}
\frac{-2\sin\phi}{w^3} & 0\\
0 & 0
\end{array}
\right ).
\end{equation}

Substituting Eq.~\eqref{C-MeanGaussian} into Eq.~\eqref{B-GP}, one obtains the geometric potential as
\renewcommand\theequation{C14}
\begin{equation}\label{C-CGP}
\tb{V}_g(\theta,r)=-\frac{\hbar^2}{8m}\frac{\sin^2\phi}{w^2}.
\end{equation}

By substituting Eqs.~\eqref{C-NVector} and ~\eqref{C-MeanGaussian} into Eq.~\eqref{B-GM}, the geometric momentum $\vec{\tb{P}}_g$ is obtained,
\renewcommand\theequation{C15}
\begin{equation}\label{C-CGM}
\vec{\tb{P}}_g=\frac{i\hbar\sin\phi}{2w}\vec{\tb{e}}_n.
\end{equation}

According to Eqs.~\eqref{ConeSurface} and ~\eqref{C-CGM}, the geometric OAM is obtained
\renewcommand\theequation{C16}
\begin{equation}\label{C-CGAM}
\vec{\tb{L}}_g=\frac{i\hbar(R\cos\phi+r)\sin\phi}{2w}\vec{\tb{e}}_{\theta}.
\end{equation}

In Cartesian coordinate system, the position vector $\vec{\tb{R}}$ Eq.~\eqref{C-Vn} can be described by
\renewcommand\theequation{C17}
\begin{equation}\label{C-CartTCurv}
\begin{cases}
x=W\cos\theta,\\
y=W\sin\theta,\\
z=r\sin\phi-q_3\cos\phi.
\end{cases}
\end{equation}
Subsequently, one can reexpress it in CCS as
\renewcommand\theequation{C18}
\begin{equation}\label{C-CurvTCart}
\begin{cases}
\theta=\arctan(\frac{y}{x}),\\
r=(\sqrt{x^2+y^2}-R)\cos\phi+z\sin\phi,\\
q_3=(\sqrt{x^2+y^2}-R)\sin\phi-z\cos\phi,
\end{cases}
\end{equation}
and then the Pauli matrices are transformed as
\renewcommand\theequation{C19}
\begin{equation}\label{PauliCurv}
\begin{split}
& \sigma^{\theta}=\frac{\partial\theta}{\partial x^i}\sigma^{i}
=\frac{1}{W} (-\sin\theta\sigma^x+\cos\theta\sigma^y),\\
& \sigma^{r}=\frac{\partial r}{\partial x^i}\sigma^i\\
&\quad =\cos\phi\cos\theta\sigma^x +\cos\phi\sin\theta\sigma^y+\sin\phi\sigma^z,\\
& \sigma^{3}=\frac{\partial q_3}{\partial x^i}\sigma^i\\
&\quad =\sin\phi\cos\theta\sigma^x +\sin\phi\sin\theta\sigma^y-\cos\phi\sigma^z,
\end{split}
\end{equation}
where $i=1,2,3$, $(\sigma^x, \sigma^y, \sigma^z)$ are the Pauli matrices. By limiting $q_3\to 0$, the reduced Pauli matrices are obtained:
\renewcommand\theequation{C20}
\begin{equation}\label{C-RPMatrix}
\begin{split}
& \sigma^{\theta}_0=\lim_{q_3\to0}\sigma^{\theta}=\frac{1}{w}(-\sin\theta\sigma^x +\cos\theta\sigma^y),\\
& \sigma^r_0=\lim_{q_3\to0}\sigma^r=\sigma^r,\quad \sigma^3_0=\lim_{q_3\to0}\sigma^3=\sigma^3.
\end{split}
\end{equation}

From Eq.~\eqref{C-CartTCurv}, one can obtain the components of the Rashba tensor $S_{\alpha\beta}$ in CCS as
\renewcommand\theequation{C21}
\begin{equation}\label{C-RTensor}
\begin{split}
& S_{\theta r}=\frac{\partial x^i}{\partial\theta}\frac{\partial x^j}{\partial r}S_{ij} =-\frac{\partial x^i}{\partial r}\frac{\partial x^j}{\partial\theta}S_{ij}=-S_{r\theta}\\
&\quad =\frac{\alpha}{\hbar}W[\sin\phi(\sin\theta +\cos\theta)-\cos\phi],\\
& S_{r3}=\frac{\partial x^i}{\partial r}\frac{\partial x^j}{\partial q_3}S_{ij} =-\frac{\partial x^i}{\partial q_3}\frac{\partial x^j}{\partial r}S_{ij}=-S_{3r}\\
&\quad =\frac{\alpha}{\hbar}(\cos\theta-\sin\theta)\\
&S_{q\theta}=\frac{\partial x^i}{\partial q_3}\frac{\partial x^j}{\partial\theta}S_{ij}=-\frac{\partial x^i}{\partial\theta}\frac{\partial x^j} {\partial q_3}S_{ij}=-S_{\theta3}\\
&\quad =\frac{\alpha}{\hbar}W[\sin\phi+\cos\phi(\sin\theta +\cos\theta)].
\end{split}
\end{equation}
By limiting $q_3\to0$, the components of the reduced Rashba tensor are obtained
\renewcommand\theequation{C22}
\begin{equation}\label{C-RRTensor}
\begin{split}
& S_{\theta r}^0=\lim_{q_3\to0}S_{\theta r}=-S_{r\theta}^0=-\lim_{q_3\to0}S_{r\theta}\\
& \quad =\frac{\alpha}{\hbar}w[\sin\phi(\sin\theta+\cos\theta) -\cos\phi],\\
& S_{r3}^0=\lim_{q_3\to0}S_{r3}=-S_{3r}^0=-\lim_{q_3\to0}S_{3r}\\
& \quad =\frac{\alpha}{\hbar}(\cos\theta-\sin\theta),\\
& S_{3\theta}^0=\lim_{q_3\to0}S_{3\theta}=-S_{\theta 3}^0=-\lim_{q_3\to0}S_{\theta 3}\\
& \quad =\frac{\alpha}{\hbar}w[\sin\phi+\cos\phi(\sin\theta +\cos\theta)].
\end{split}
\end{equation}
Similarly, the components of the Dresselhaus tensor $S_{\alpha\alpha\beta\beta}$ can be expressed as
\renewcommand\theequation{C23}
\begin{equation}\label{C-DTensor}
\begin{split}
&S_{\theta\theta r r}=\frac{\partial x^i}{\partial\theta} \frac{\partial x^i}{\partial\theta}\frac{\partial x^j}{\partial r} \frac{\partial x^j}{\partial r}S_{iijj}=-\frac{\partial x^i}{\partial r} \frac{\partial x^i}{\partial r}\frac{\partial x^j}{\partial\theta} \frac{\partial x^j}{\partial\theta}S_{iijj}\\
&\quad\quad =-S_{rr\theta\theta}=-\frac{\beta}{\hbar^3} W^2\cos2\phi\cos2\theta\\
& S_{rr33}=\frac{\partial x^i}{\partial r}\frac{\partial x^i}{\partial r} \frac{\partial x^j}{\partial q_3}\frac{\partial x^j}{\partial q_3}S_{iijj} =-\frac{\partial x^i}{\partial q_3}\frac{\partial x^i}{\partial q_3} \frac{\partial x^j}{\partial r}\frac{\partial x^j}{\partial r}S_{iijj}\\
&\quad\quad =-S_{33rr}=-\frac{\beta}{\hbar^3}\cos2\phi\cos2\theta,\\
& S_{33\theta\theta}=\frac{\partial x^i}{\partial q_3} \frac{\partial x^i}{\partial q_3}\frac{\partial x^j}{\partial\theta} \frac{\partial x^j}{\partial\theta}S_{iijj}=-\frac{\partial x^i}{\partial\theta} \frac{\partial x^i}{\partial\theta} \frac{\partial x^j}{\partial q_3} \frac{\partial x^j}{\partial q_3}\\
&\quad\quad =-S_{\theta\theta33} =-\frac{\beta}{\hbar^3}W^2\cos2\phi\cos2\theta.
\end{split}
\end{equation}
By limiting $q_3\to0$, the components of the reduced Dresselhaus tensor are obtained
\renewcommand\theequation{C24}
\begin{equation}\label{C-RDTensor}
\begin{split}
& S_{\theta\theta rr}^0=\lim_{q_3\to0}S_{\theta\theta rr}=-S_{rr\theta\theta}^0=-\lim_{q_3\to0}S_{rr\theta\theta}\\
& \quad =-\frac{\beta}{\hbar^3}w^2\cos2\phi\cos2\theta,\\
& S_{rr33}^0=\lim_{q_3\to0}S_{rr33}=-S_{33rr}^0=-\lim_{q_3\to0}S_{33rr}\\
& \quad =-\frac{\beta}{\hbar^3}\cos2\phi\cos2\theta,\\
& S_{33\theta\theta}^0=\lim_{q_3\to0}S_{33\theta\theta} =-S_{\theta\theta33}^0=-\lim_{q_3\to0}S_{\theta\theta33}\\
& \quad =-\frac{\beta}{\hbar^3}w^2\cos2\phi\cos2\theta.
\end{split}
\end{equation}

\end{document}